\documentclass[conference]{IEEEtran}
\IEEEoverridecommandlockouts

\usepackage{amsmath,amssymb,amsfonts}
\usepackage{cite}
\usepackage{algorithm}
\usepackage{graphicx}
\usepackage{xcolor}
\usepackage{booktabs}
\usepackage{multirow}
\usepackage{float}
\usepackage{array}
\usepackage{multirow}  

\def\BibTeX{{\rm B\kern-.05em{\sc i\kern-.025em b}%
\kern-.08em T\kern-.1667em\lower.7ex\hbox{E}\kern-.125emX}}

\begin{document}

\title{Real-time Load Current Monitoring of Overhead Lines Using GMR Sensors

\thanks{This work was financially supported in part by the New Zealand Science for Technological Innovation National Science Challenge (contract RTVU1702); Ministry of Business, Innovation and Employment, New Zealand (contract RTVU1811), Victoria University of Wellington Research Trust, and the Office of Research and Sponsored Programs Administration, Lamar University. The authors are with \textsuperscript{1}Department of Electrical and Computer Engineering, Lamar University, Beaumont, TX, USA, and \textsuperscript{2}Robinson Research Institute, Gracefield, New Zealand. Corresponding Author: Fiona J. Stevens McFadden, fiona.stevensmcfadden@vuw.ac.nz}}

\author{
Md Mahfuzur Rahman Chy\textsuperscript{1}, Md Rifat Al Amin Khan\textsuperscript{1}, Md Sultan Mahamud\textsuperscript{1}, \\
Anwarul Islam Sifat\textsuperscript{1}, Fiona J. Stevens McFadden\textsuperscript{2}
 
}
\maketitle
\begin{abstract}
Non-contact current monitoring has emerged as a prominent research focus owing to its non-intrusive characteristics and low maintenance requirements. However, while they offer high sensitivity, contactless sensors necessitate sophisticated design methodologies and thorough experimental validation. In this study, a Giant Magneto-Resistance (GMR) sensor is employed to monitor the instantaneous currents of a three-phase 400-volt overhead line, and its performance is evaluated against that of a conventional contact-based Hall effect sensor. A mathematical framework is developed to calculate current from the measured magnetic field signals. Furthermore, a MATLAB-based dashboard is implemented to enable real-time visualization of current measurements from both sensors under linear and non-linear load conditions. The GMR current sensor achieved a relative accuracy of 64.64\%–91.49\%, with most phases above 80\%. Identified improvements over this are possible, indicating that the sensing method has potential as a basis for calculating phase currents.

\end{abstract}

\begin{IEEEkeywords}
GMR, Test Facility, MATLAB, Overhead Lines, Hall-effect Sensor
\end{IEEEkeywords}

\section{Introduction}

Current monitoring schemes for distribution lines require safe and cost-friendly solutions. Existing methods for measuring overhead line currents usually require direct contact with the conductors, which can be costly, hazardous to personnel, and susceptible to electromagnetic interference. 

Typically, the Hall effect sensor, Rogowski coil, shunt resistor, and current transformer (CT) are the most common current sensors \cite{yang2024prospective, swain2022sensor}. CT and Rogowski coils can only measure alternating current, and CTs are subject to saturation \cite{ferreira2010noninvasive}. In contrast, Rogowski coils typically exhibit low sensitivity due to their wide measurement bandwidth and are not immune to environmental and system interference \cite{lu2024parametric}. Magnetic sensors such as anisotropic magneto-resistance (AMR), GMR, colossal magneto-resistance (CMR), and tunnel magneto-resistance (TMR) have recently gained popularity. Their inherent physical design, for example, tuning the bias voltage, allows them to have higher measurement bandwidth \cite{dkabek2016sensitivity}. The key advantages of these sensors are electrical safety through galvanic isolation, low cost, wide measurement bandwidth, and immunity to electromagnetic noise.

The AMR sensor uses a permalloy element whose resistance changes when the conductor’s magnetic field alters its internal magnetization \cite{bernieri2017amr}. The narrow linear operating range makes it prone to magnetic saturation, which restricts the achievement of high signal-to-noise ratios \cite{sun2024high}. CMR technology concentrates the magnetic field onto a manganite sensor using a flux concentrator \cite{vsimkevivcius2013current}. The sensor's resistance changes in response to the magnetic field. This resistance change is then measured and converted to a corresponding current value. However, it requires high magnetic fields to induce measurable resistance variations, and the complexity of its fabrication process further limits its practicality for real-world applications \cite{nickel1995magnetoresistance, balevivcius2024magnetic}. TMR sensors exhibit high sensitivity to sensing axis orientation. They are also more immune to electromagnetic noise, have a long linear range, and maintain stability across a broader temperature range compared to other magnetoresistive sensors \cite{chen2022intelligent}. However, the experimental findings suggest that for low current sensing ($\leq$2A), the TMR sensor exhibits approximately twice the error compared to the GMR sensor \cite{shrawane2020performance}. In addition, the commercially available GMR sensor is significantly more cost effective compared to TMR and AMR sensor \cite{khan2021magnetic}. 

The Giant Magneto-Resistance sensor is a highly sensitive device used to detect magnetic fields. It operates based on the giant magneto-resistance effect \cite{djamal2012development}. When an external magnetic field is applied, the alignment of the magnetic moments in the ferromagnetic layers changes, leading to a significant change in the sensor's electrical resistance \cite{reig2013giant}.

Current measurement using non-contact sensors has been extensively studied in previous works. In \cite{ouyang2019current}, multiple GMR sensors are placed in a magnetic ring configuration to optimize field concentration while minimizing measurement errors and electromagnetic noise. Subsequent developments addressed a matrix-based decoupling approach \cite{olson2010effective} to calculate the currents in multiple current-carrying conductors that produce a cross-coupled magnetic flux. Using inverted geometric relationships between sensors and conductors, individual conductor currents are decoupled. In \cite{alvi20192}, GMR sensors based on point field detectors (PFDs) can calculate inverter current by analyzing the superimposed two-dimensional magnetic field. In \cite{zhang2019current}, a 3D TMR-sensor-based ring-type array field detector is developed to address measurement errors caused by inclined conductors that are not perpendicular to the sensor. Overall, these approaches were limited to the use of non-contact sensors in close proximity to the current-carrying conductor, thereby restricting their applicability in measuring the current of overhead lines at a distance. 

In this paper, we leverage off-the-shelf GMR sensors to measure the magnetic fields in overhead lines at a distance and use these measurements to calculate the phase currents.
Additionally, we have developed a fully functional MATLAB-powered application dashboard that provides real-time data visualization of alternating currents passing through the overhead lines, comparing the calculated currents from non-contact magnetic sensors with those from contact-based Hall-effect current sensors.




\section{Methodology}
In an overhead line setup, each conductor generates a magnetic field that couples with the fields of the other conductors. The magnetic field detected by a single sensor is influenced by the spacing between conductors, and a single 2D GMR sensor-head measurement alone cannot fully decouple these magnetic fields  \cite{khawaja2017estimation}. In our experimental setup, a geometric relationship is established between two GMR sensor heads and the locations of the overhead conductor at the $x$ (conductor position) and $y$ (phase current flow direction) planes, yielding magnetic field output coefficients that were also reported in our previous work \cite{sifat2020characterization}. The horizontal $B_{x}$ and vertical $B_{z}$ components of the magnetic field at the sensor head are formulated using integral expressions derived from the Biot–Savart law. These integrals compute the total magnetic field contribution by summing the effects of differential current elements distributed along the span of each overhead conductor. 
For brevity, the resulting field expressions are encapsulated by spatial coefficients $\mathrm{C_{x}}$ and $\mathrm{C_{z}}$, leading to:


\begin{equation}\label{BxBz1}
\begin{aligned}
\mbox{\fontsize{10}{8}\selectfont\(\vec{B_x} = 
\frac{\mu_0I}{4\pi} \mathrm{C_x} \)}, 
\mbox{\fontsize{10}{8}\selectfont\(\vec{B_z} = 
\frac{\mu_0I}{4\pi} \mathrm{C_z} \)}
\end{aligned}
\end{equation}

The positions of the overhead conductors are defined by 
setting the value of $x^{'}_{0}$ w.r.t the horizontal plane $x$ axis. Therefore, 
the coefficients $\mathrm{C_x}$ and $\mathrm{C_z}$ are a function of 
$x^{'}_{0}$, i.e. $\mathrm{C_x}(x^{'}_{0})$ and $\mathrm{C_z}(x^{'}_{0})$. Let $k_1$ and $k_2$ be the positions of two 2D sensors under the overhead lines. The final magnetic flux density equations of the 
\textit{x}- and \textit{z}-axes for each of the overhead line conductors can 
be presented using (\ref{Bx}) and (\ref{Bz}).  
\begin{equation}
\resizebox{0.4\textwidth}{!}{$
\begin{aligned}
\vec{B_x}(\mathrm{A, B, C, N}) = \frac{\mu_0 I}{4\pi} \bigg[
&\mathrm{C_x}(x'_0(k_1, k_2)) + 
\mathrm{C_x}(x'_0(k_1, k_2)) + \\
&\mathrm{C_x}(x'_0(k_1, k_2)) + 
\mathrm{C_x}(x'_0(k_1, k_2)) 
\bigg]
\end{aligned}
$}
\label{Bx}
\end{equation}
\vspace{-1em}
\begin{equation}
\label{Bz}
\resizebox{0.4\textwidth}{!}{$
\begin{aligned}
\vec{B_z}(\mathrm{A, B, C, N}) = \frac{\mu_0 I}{4\pi} \bigg[
&\mathrm{C_z}(x'_0(k_1, k_2)) +
 \mathrm{C_z}(x'_0(k_1, k_2)) + \\
&\mathrm{C_z}(x'_0(k_1, k_2)) +
 \mathrm{C_z}(x'_0(k_1, k_2))
\bigg]
\end{aligned}
$}
\end{equation}

\begin{figure}
    \centering
    \includegraphics[width=0.25\textwidth]{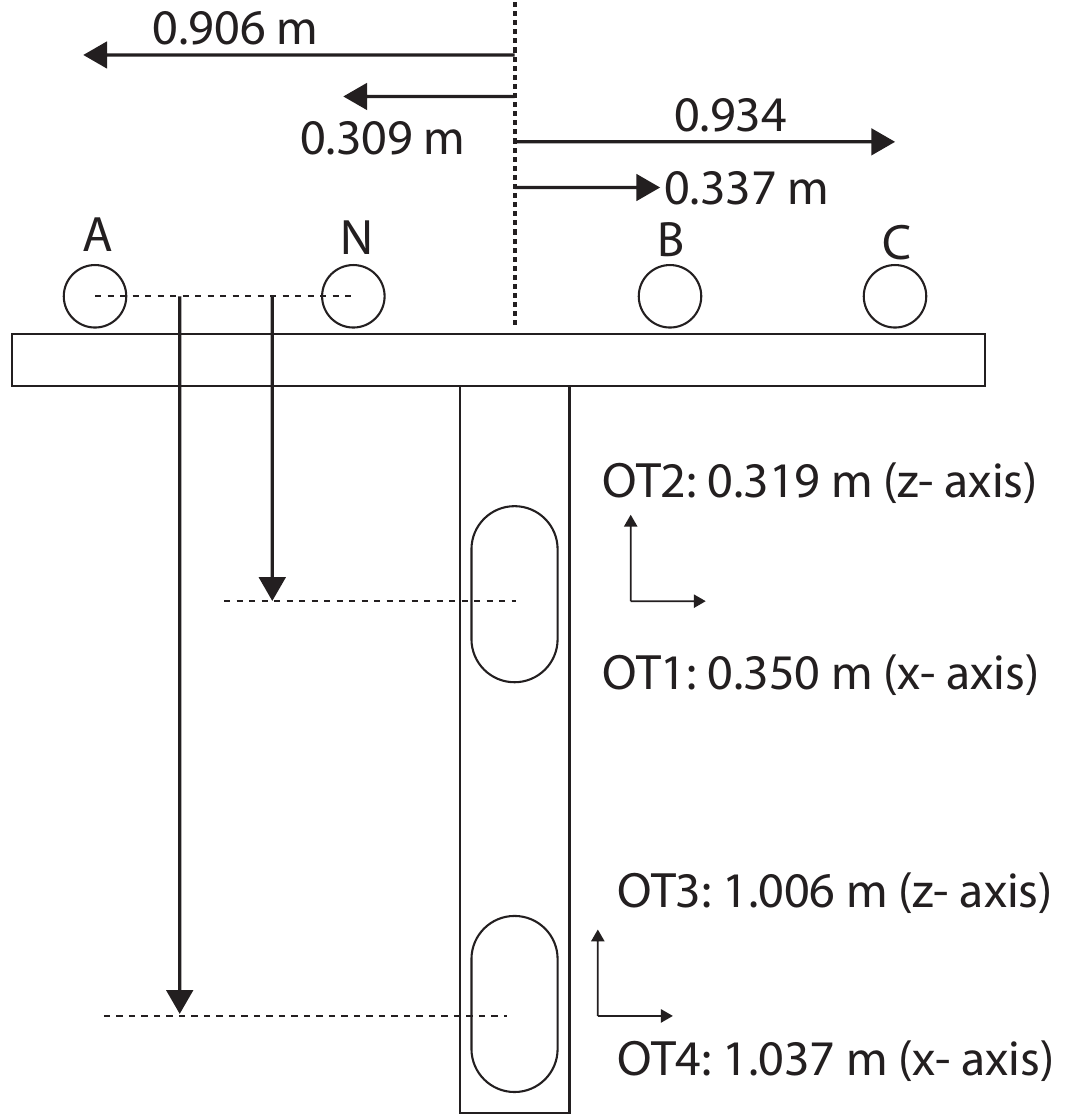}
    \caption{The position of the GMR sensor with respect to the overhead power lines is illustrated. Here, OTn (where n = 1 to 4) denotes the single-axis GMR sensor mounted on a 3D-printed sensor holder.}
    \label{fig: Pole_config}
    \vspace{-1em}
\end{figure}

For each of the conductors, (\ref{Bx}), and (\ref{Bz}) will form a $2 
\times 1$ vector in the \textit{x}- and \textit{z}-axes direction. The $\mathrm{C_x}$ and $\mathrm{C_z}$ vectors can be concatenated together to 
build a cross-coupled matrix $\mathrm{C_{xz}}$ of size $ 4 \times 4$. 
The cross-coupled matrix consists of sub-vectors in each column that represent the coefficients for the \textit{$x$} and \textit{$z$} components of each overhead line. The dimensions of this matrix are determined by the number of sensor heads. The values of the coefficients are computed based on the vertical  placement of the sensors beneath the overhead lines (Fig. \ref{fig: Pole_config}).


The cross-coupled matrix is geometry-dependent and thus remains constant based on the relative placement of the sensor heads. In practical applications, the matrix is computed after installing the sensor head on the pole. The field vectors in the \textit{x} and \textit{z} direction are multiplied by the inverted cross-coupled matrix to derive the current $I$. Therefore, (\ref{BxBz1}) can be re-written as,  
\begin{equation}\label{eq:I-back}
\resizebox{0.23\textwidth}{!}{$
I = \frac{4\pi}{\mu_0} \ \mathrm{C_{xz}}^{-1}
\left[
\begin{aligned}
\vec{B_x}(\mathrm{A,B,C,N})\\
\vec{B_z}(\mathrm{A,B,C,N})
\end{aligned}
\right]
$}
\end{equation}
Here, $\vec{B_x}(\mathrm{A,B,C,N})$ and $\vec{B_z}(\mathrm{A,B,C,N})$ are the 
cross-coupled magnetic fields measured by the sensor heads. Each row of the 
output matrix of size (4 $\times$ Number of samples (Ns)) in (\ref{eq:I-back}) 
represents the calculated current of the overhead lines. 

\subsection{GMR Sensor-based Real-Time Load Current Sensing application in MATLAB}

\begin{figure*}
    \vspace{-1em}
    \centering
    \includegraphics[width=0.9\textwidth]{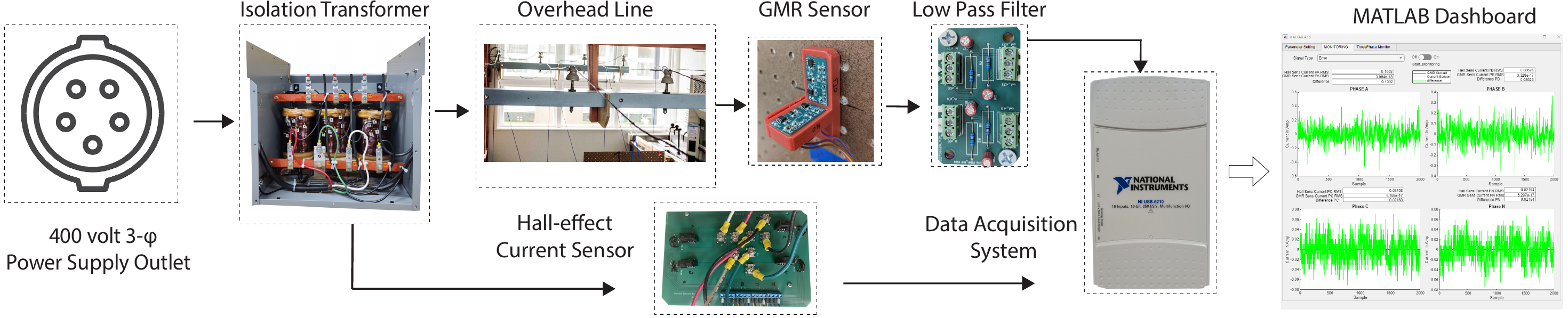}   
    \caption{A three-phase isolated delta-wye configured transformer supplies stable power to the load through overhead lines. The Hall-effect and GMR sensors are powered by five and eight-volt regulated DC power supplies. Each GMR sensor output is filtered via a low-pass filter to eliminate the anti-aliasing effect and high-frequency noise. The Hall effect sensors have integrated filter units; therefore, external filters are unnecessary. Both types of sensor data are subsequently sent to the National Instruments Data Acquisition System (NI-DAQ). Finally, the data from the NI-DAQ is processed and plotted using MATLAB application.}
    \label{fig:current_sensing_system}
    \vspace{-1em}
\end{figure*}

A MATLAB-powered application dashboard is developed, in which an user can select the desired signal for monitoring via a drop-down menu (Fig. \ref{fig:current_sensing_system}). The menu offers seven signal options. Calculated Current displays phase current derived from GMR sensor magnetic-field data via the field-to-current transformation. Measured Current shows the phase current from the Hall-effect sensor. Measured and Calculated Currents presents both datasets side by side for direct comparison and evaluation of the GMR-based calculation method.

Within the MATLAB script, magnetic field data are acquired from an array of four GMR sensors using an NI-DAQ system. The DAQ records the raw analog voltage output from each sensor channel, which is subsequently converted into magnetic-field components by multiplying the voltages by the sensor's calibrated conversion factor as defined in (Section \ref{Experimentation and Result}). The program then defines the geometric and configuration parameters of the measurement setup, including the vertical positions of the sensor configuration, the span length, and the spatial positions of the four phase conductors (A, B, C, and N). Fixed sag lengths for each phase are also specified, along with a fixed span discretisation step size. All these values are constants  $\mathcal{O}(1)$ determined by the experimental arrangement and remain unchanged throughout execution. For each incoming DAQ sample, i.e., magnetic field vector, the precomputed 4×4 inverse C-matrix is multiplied in real time to yield the corresponding current vector. This constant-time operation is repeated for all $\mathrm{Ns}$ samples, resulting in an overall computational complexity of $\mathcal{O}(\mathrm{Ns})$. Plotting measured versus calculated fields and currents, and saving results, also iterate over $\mathrm{Ns}$-length arrays, so the total runtime is $\mathcal{T}(\mathrm{Ns}) = \mathcal{O}(1)  + \mathcal{O}(\mathrm{Ns})$. For large Ns, the constant term is negligible, leaving the runtime dominated by the linear term $\mathcal{O}(\mathrm{Ns})$.

We profiled MATLAB application execution time to evaluate the time it takes to process, that is, calculate currents from magnetic fields and plot one complete cycle of calculated current waveforms. At the maximum tested sampling rate of 28 kHz, processing takes 0.0034 seconds (s) to calculate the currents, and the four plots take a total of 0.1374s to render. To assess real-time scalability, the application is tested at sampling rates ranging from 1 kHz to 28 kHz. The results indicate that execution time exhibits approximately linear growth with increasing sampling rates, attributable to the additional computational overhead inherent in the application’s processing routines.

\section{Experimentation and Result}
\label{Experimentation and Result}
Our previous work \cite{sifat2019facility}, which utilized GMR sensors, focused on detecting high-impedance faults. This present study focuses on monitoring the line current under no-fault test conditions. Therefore, the facility has four principal components: the power section, overhead line setup, sensor design, calibrated GMR sensors, and load section (Fig. \ref{fig:current_sensing_system}). 

The power section starts by isolating a three-phase alternating current (AC) power supply from the test facility using a three-phase delta-wye configured 12 kVA, 415/415V, 1:1 turn ratio, 50 Hz isolation transformer. The transformer output terminal is connected to a 6m long overhead line's spatial layout. The spatial layout employs four conductors, including the neutral phase conductor. The arrangement follows a catenary profile, replicating realistic sag conditions. The overhead lines are installed in the lab with a single span length of 3m.

The test facility uses Hall-effect and GMR sensors for current measurement. Four Hall effect sensors, each connected in series with a conductor, are installed on a PCB, allowing direct current measurement \cite{crescentini2021hall}. In our experimental study, a HO 25-P type Hall effect sensor is used, with a sensitivity of 32 mV/A. In contrast, the GMR sensor measures changes in the magnetic field surrounding the overhead line setup. Then, the magnetic field is converted to calculate phase currents. A GMR sensor is a single-axis device that detects magnetic field variations on only one coordinate axis. As part of our test configuration, two GMR sensors were mounted on two PCB boards to form a single GMR sensor head, enablig simultaneous recording of magnetic fields from both the x and z axes.

The 3D-printed magnetic sensor head was constructed from polylactic acid (PLA) thermoplastic. It houses the GMR sensor along with its peripheral connections. The sensor head is mounted on a PCB board and features an "L"- shaped design with two designated placeholders. The z-axis sensor PCB is positioned on the vertical placeholder, while the x-axis sensor PCB is placed on the horizontal placeholder, aligning parallel to the overhead line arrangement.


Typically, the output of a GMR sensor at the maximum applied field is 40 mV/V. This low-voltage output is amplified using an instrumentation operational amplifier (op-amp). Each sensor in the sensor head had a dedicated op-amp to amplify the signal from the GMR sensor. The voltage output of the sensor $\mathrm{V_{sns}}$ is converted to magnetic flux density by multiplying it with the magnetic flux density factor, $\mathrm{MFDF = 1 / (((V_{\text{op}} \times S_{\text{ns}}) / 100) \times 10^3)}$. Here, $\mathrm{S_{ns}}$ (V/V-Oe) is the GMR sensitivity factor, $\mathrm{V_{sns}}$ and $\mathrm{V_{op}}$ (V) are the sensor output voltage and the operating voltage.

The typical sensitivity range of AAH002-02 GMR sensor \cite{nve} is 11–18 mV/V-Oe. Therefore, we need to individually calibrate the four sensors installed on each of the two sensor heads. We calibrated the sensor using a solenoid coil that generates a magnetic field around a closed path, applying currents ranging from -100 to 100 mA with a current source meter. The sensors were placed inside a solenoid to capture the magnetic field generated by the coil. We estimated the slope of the linear region from the sensor characteristic curve, which plots the applied magnetic field against the output voltage of a GMR sensor, to determine the sensitivity range. We use the mean positive sensitivity slope value to calculate the MFDF.

We use both linear and nonlinear loads to generate load currents. Linear loads included twin 2 kW halogen lamps, a 1 kW resistive heater, and a 1 kW air conditioner. Nonlinear loads consisted of a microwave oven, a 1 kW air conditioner with two electronic compact fluorescent lamps (CFLs) banks, totaling 230 W, and a vacuum cleaner, which introduced current waveform distortion compared to linear loads.

\subsection{Measured and calculated load current}

\begin{figure}[!ht]
     \vspace{-1em}
    \centering
    \includegraphics[width=0.49\textwidth]{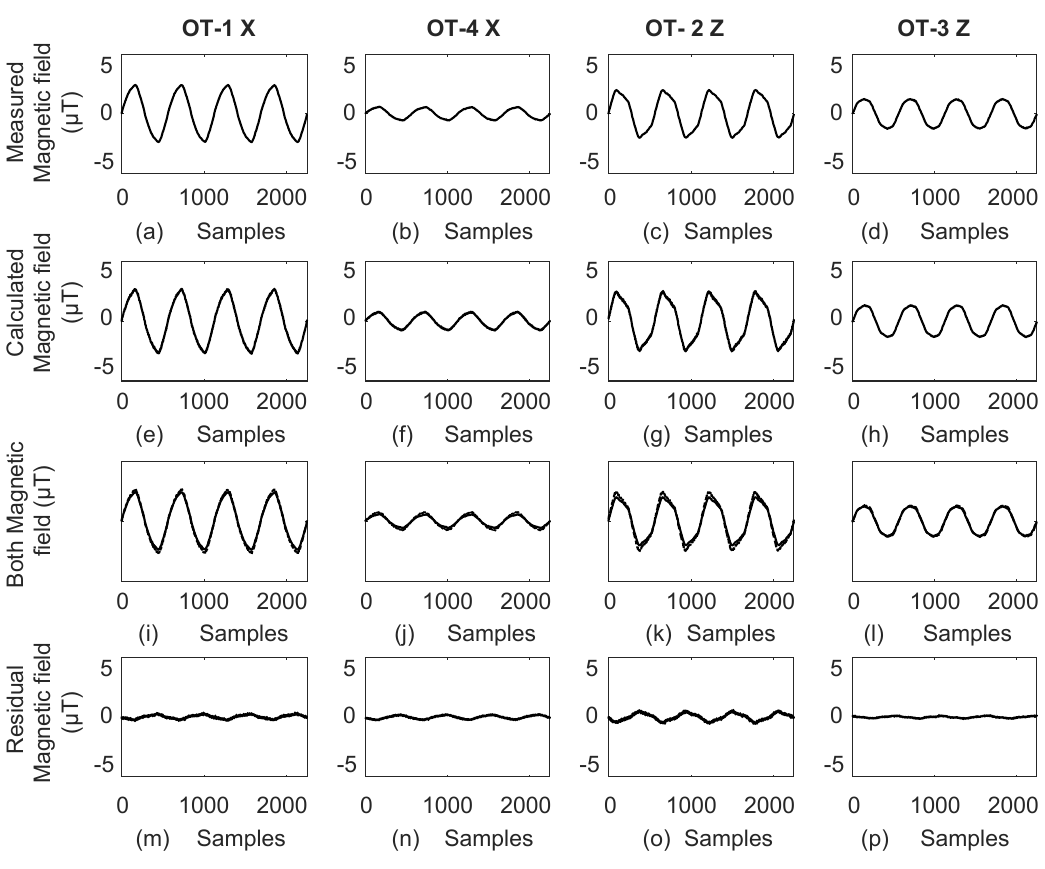}
    \caption{
    Comparative analysis under linear load conditions shows that the measured magnetic field from the GMR sensor closely overlaps with the field calculated using (\ref{BxBz1}) (3rd row). The residual field (4th row) represents the instantaneous difference between the measured and calculated values.} 
    \label{Result_magnetic_L}
    \vspace{-1em}
\end{figure}

\begin{figure}[!ht]
    \centering
    \includegraphics[width=0.49\textwidth]{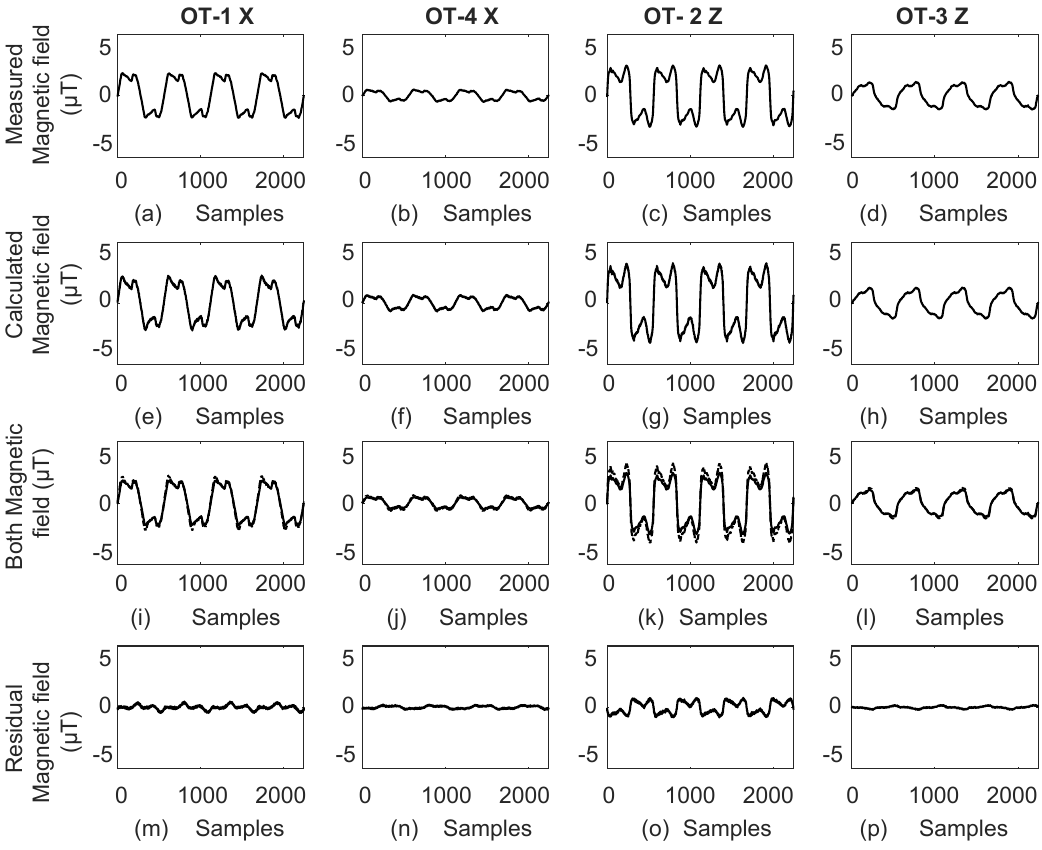}
    \caption{
    Comparative analysis under non-linear load conditions shows that the measured magnetic field from the GMR sensor closely overlaps with the field calculated using (\ref{BxBz1}) (3rd row). The residual field (4th row) represents the instantaneous difference between the measured and calculated values.} 
    \label{Result_magnetic_NL}
\end{figure}

\begin{figure}[!ht]
    \centering
    \includegraphics[width=0.49\textwidth]{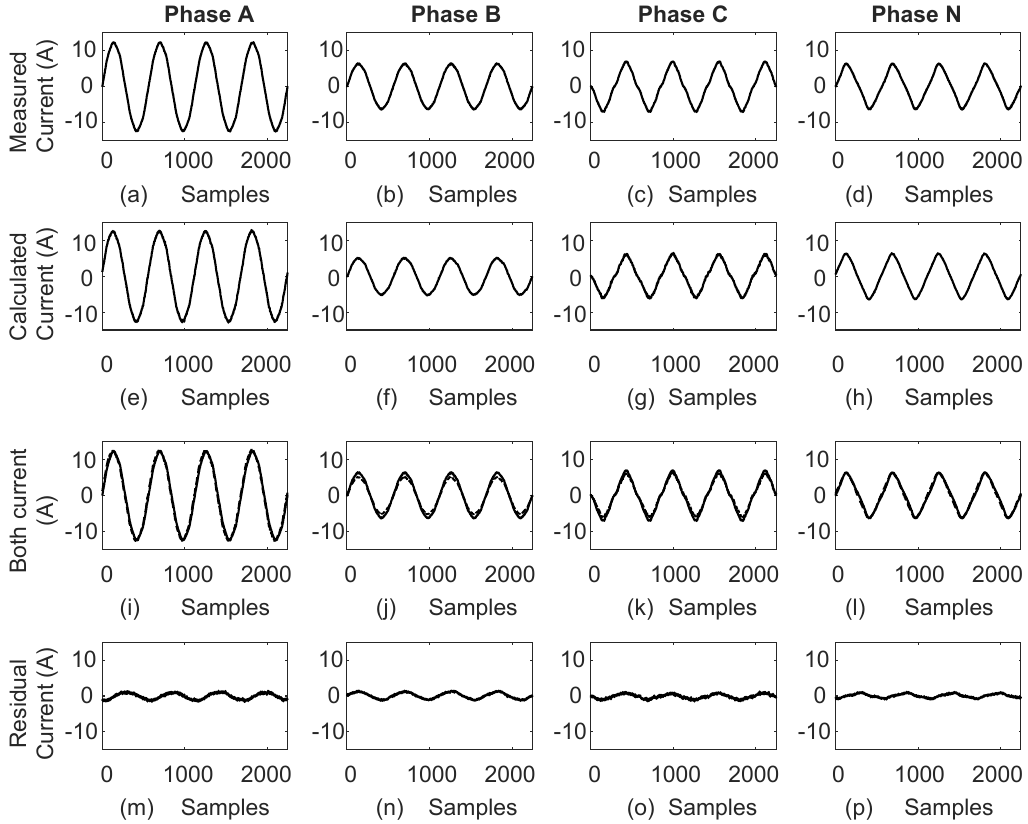}
    \caption{
    Comparative analysis under linear load conditions shows that the phase current calculated from GMR sensor data using (\ref{eq:I-back}) closely matches the measured current signal (3rd row). The residual current remains within $\pm 2\,\mathrm{A}$ (4th row).}
    \label{Result_current_L}
     \vspace{-1em}
\end{figure}

\begin{figure}[!ht]
    \centering
    \includegraphics[width=0.49\textwidth]{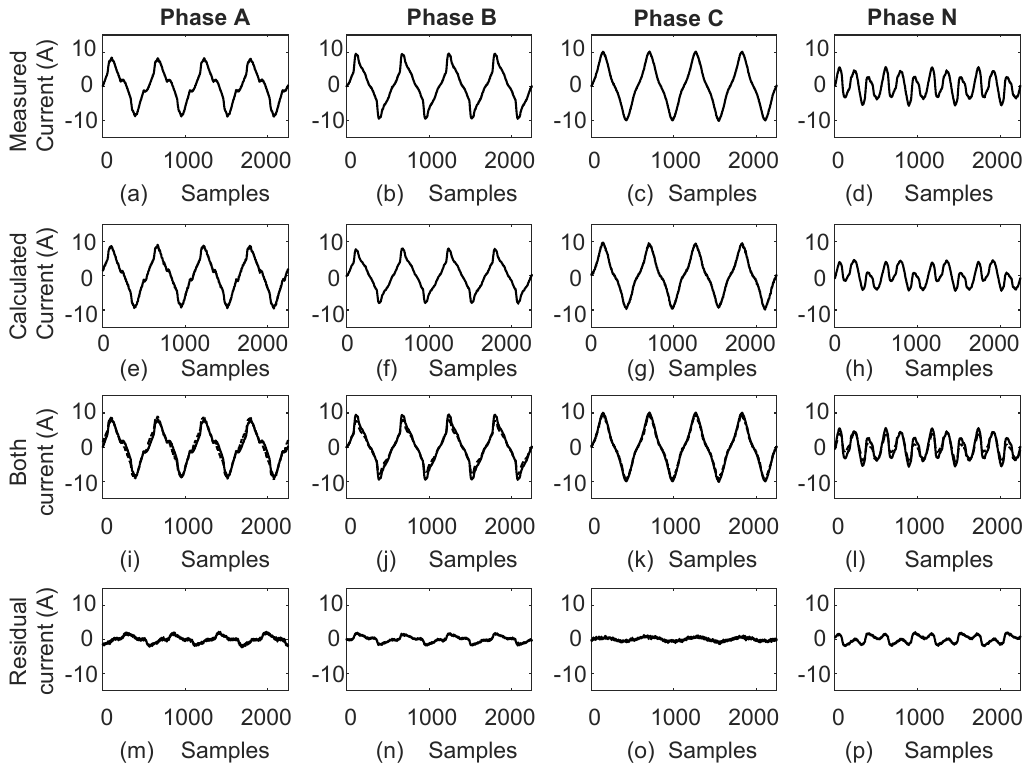}
    \caption{Comparative analysis under linear load conditions shows that the phase current calculated from GMR sensor data using (\ref{eq:I-back}) closely matches the measured current signal (3rd row). The residual current remains within $\pm 2\,\mathrm{A}$ (4th row).}
    \label{Result_current_NL}
    \vspace{-1em}
\end{figure}

We compared the measured and calculated magnetic fields and load currents from the GMR and Hall effect sensors. While the GMR sensor detects the alternating magnetic field generated by the overhead lines, the field can also be calculated using (\ref{Bx}). The measured field data from an array of GMR sensors is multiplied by an inverse coefficient matrix to calculate the currents using (\ref{eq:I-back}). We calculated the difference between the measured and calculated instantaneous currents, i.e., the residual current, and the normalized mean absolute error (NMAE) to assess the accuracy of GMR as a current sensor under normal (linear) and harmonic (nonlinear) load conditions. 

Figs. \ref{Result_magnetic_L} and \ref{Result_magnetic_NL} depict the alternating magnetic fields generated from linear and nonlinear loads. The top OT-1 and OT-2 sensors recorded stronger magnetic fields due to their closer proximity to the overhead lines than the bottom OT-3 and OT-4 sensors. 

The calculated currents for linear and non-linear load conditions are shown in Figs. \ref{Result_current_L} and  \ref{Result_current_NL}. The distorted waveform in the neutral phase is due to the presence of odd-order harmonics in the non-linear loads. The NMAE values in Table \ref{tab:RMSchart} represent the mean absolute deviation between the calculated and measured currents, scaled by the mean of the measured current magnitudes, to provide a dimensionless, scale-independent error metric. Under linear load conditions, the calculated current of phase B has the highest NMAE of 18.36\% compared to the measured Hall-effect current. Under non-linear load conditions, the largest deviation is observed in Phase N, with an NMAE of 35.36\%, where the increased harmonic distortion in the phase currents (Fig. \ref{Result_current_NL} (l)) makes accurate current calculation more challenging. The source of error could be sensor placement with respect to overhead lines, the orientation of the sensing axis with respect to true horizontal and vertical, and/or hysteresis error in the sensor. The directional sensitivity of the GMR sensor makes it susceptible to measurement deviations caused by slight misalignments.

\begin{table}
  \centering
  \caption{Comparison of measured and calculated root mean square (RMS) current, and their NMAE}
    \begin{tabular}{llccc}
    \toprule
    \multicolumn{1}{c}{\textbf{Load}} & \multicolumn{1}{c}{\textbf{Phase }} & \textbf{Measured } & \textbf{Calculated} & \textbf{Normalized} \\
    \multicolumn{1}{c}{\textbf{type}} &       & \textbf{Current (A)} & \textbf{Current (A)} & \textbf{MAE (\%)} \\
    \midrule
          & A     & 8.649 & 8.832 & 9.46 \\
    Linear  & B     & 4.403 & 3.597 & 18.36 \\
    Load  & C     & 4.436 & 3.856 & 14.36 \\
          & N     & 4.046 & 4.120 & 12.47 \\
    \midrule
          & A     & 4.751 & 5.166 & 19.47 \\
    \multicolumn{1}{p{4.555em}}{Non-linear } & B     & 5.089 & 4.263 & 17.02 \\
    \multicolumn{1}{p{4.555em}}{Load} & C     & 5.935 & 5.515 & 8.51 \\
          & N     & 3.047 & 2.587 & 35.36 \\
    \bottomrule
    \end{tabular}%
  \label{tab:RMSchart}%
  \vspace{-1.5em}
\end{table}%
\vspace{-0.3em}
\section{Conclusion}
We calculated the phase currents at a distance from the distribution overhead lines using a vertically positioned array of GMR sensors. We validated the GMR sensor's current-sensing capability using a purpose-built test facility that accurately represents low-voltage overhead distribution power lines. Two calibrated GMR sensor heads measured the magnetic field, with sensitivity factors converting sensor voltages to magnetic field values. Hall-effect sensors provided reference current measurements, while a modified Biot-Savart law-based model calculated line currents from the GMR data. To assess the accuracy of GMR sensors, the calculated value was then compared with the current measured by the Hall effect sensor under linear and nonlinear load conditions. Finally, a MATLAB-based application was developed to monitor, in real time, the measured and calculated magnetic field and load current data. 

Future work will prioritize the development of a systematic method for post-assembly calibration of the sensor head modules under AC conditions, followed by a method to incorporate misalignment parameters into magnetic field calculations to minimize calculation error.


\bibliographystyle{IEEEtran}
\bibliography{bibliography}

\end{document}